\documentclass[twocolumn,aps,prx]{revtex4}

\usepackage{graphics}% Include figure files
\usepackage{graphicx}% Include figure files
\usepackage{dcolumn}% Align table columns on decimal point
\usepackage{bm}% bold math
\usepackage{amssymb,amsmath,graphicx,graphics,color,epstopdf}

\begin{document}

%\draft

\title{Global Hawking Temperature of Schwarzschild-de Sitter Spacetime: a Topological Approach}
\author{Charles W. Robson, Leone Di Mauro Villari and Fabio Biancalana}
\affiliation{School of Engineering and Physical Sciences, Heriot-Watt University, EH14 4AS Edinburgh, UK}
\date{\today}

\begin{abstract}

We introduce a calculation, based on purely topological reasoning, of the global equilibrium Hawking temperature of the Schwarzschild-de Sitter spacetime, where a Schwarzschild black hole horizon coexists with a de Sitter cosmological horizon. Our method is based on the careful calculation of the Euler characteristic of the total system, showing that this quantity completely determines the thermodynamical features of the system. The method is universal and can be applied to any structure possessing multiple horizons in general relativity.

\end{abstract}

\maketitle

\section{Introduction}

The thermodynamical properties of black holes are among the most interesting subjects in modern physics, compelling not only in the context of astrophysics but also due to the mixture of information theory, general relativity and quantum field theory needed in order to study these extreme objects. Probing these systems is surely a promising route towards a theory of quantum gravity. The discovery that the entropy of a black hole is proportional to its area has also stimulated the emergence of theories proposing geometrical information storage for the Universe itself \cite{Bekenstein,Susskind,Susskind2,Bousso3}.

Over the past few decades, since Hawking's original discovery that black holes emit radiation, a massive amount of effort has been invested in scrutinising the thermodynamics of black holes \cite{Hawking_rad}. Many black hole systems are now well understood, even in spacetimes with number of dimensions other than four \cite{Chandra,Lemos}. One system yet to be fully understood is the de Sitter universe (with a positive cosmological constant $\Lambda>0$) containing a Schwarzschild black hole, having proven especially difficult to deal with \cite{Hawking2,Pappas,Kubi}. This difficulty arises as two horizons are present in the spacetime, the event horizon of the black hole and a cosmological horizon. Defining a single new equilibrium Hawking temperature for the entire spacetime, taking into account both horizons and using only topological quantities, is the novel result of this work.

An early work studying the thermodynamics of Schwarzschild-de Sitter spacetime by Gibbons and Hawking elucidated one of the main challenges, namely that of covering the whole spacetime with an appropriate coordinate system \cite{Hawking2} . Covering two horizons (with surface gravities for example of $\kappa_{1}$ and $\kappa_{2}$) with a single Kruskal-like set of coordinates leads to problems, in their words: ``[The thermodynamical] difficulties arise from the fact that when one has two or more sets of horizons with different surface gravities one has to introduce separate Kruskal-type coordinate patches to cover each set of horizons ... There is thus a branch cut in the relation between the two coordinate patches if $\kappa_{1}\neq \kappa_{2}$". We believe that the approach introduced in our work has a distinct advantage over coordinate-based studies of this spacetime as we define a Hawking temperature for the spacetime \emph{topologically}.

It has been known for several years that black holes' entropies can be studied using topological methods however only recently has it emerged in work by the authors that the Hawking temperature of black hole spacetimes can be studied in a purely topological manner as well \cite{Gibbons,Liberati,ourpaper2}. In our previous work, several known black hole systems were studied using a topological approach, based upon an invariant known as the Euler characteristic, which was shown to produce the correct, accepted Hawking temperature of black holes in every case. In this work we push our topological approach further, showing that it can be used to study \emph{cosmological horizons} in de Sitter (dS) spacetime and by extension can also define a global topological Hawking temperature for the Schwarzschild-de Sitter (SdS) spacetime in thermal equilibrium. We believe this is the first topological derivation of a Hawking temperature of SdS spacetime yet found.

Defining temperature(s) for SdS spacetime is a much-debated topic in the literature. Many different temperatures have been suggested: generally a distinct one for each horizon, but in some cases an effective one formed from a combination of each horizon temperature, and sometimes one measurable only by a specially-positioned observer \cite{Pappas}. In this work we define a global SdS temperature using a topological technique, suggesting that the entire spacetime, including both horizons, can exist in a thermal equilibrium. Our approach seems more natural as a basis for defining an SdS spacetime temperature than previous attempts as it relies on no thermodynamical assumptions, is not sensitive to any complications due to coordinate system choice, and is based purely on the fixed points of Killing vector fields defined on the spacetime.

Our result supports the assertion that the Schwarzschild black hole in the de Sitter background acts like a black hole inside a finite radiation-filled box; see \cite{Shanki}. If the box is bounded then the system will be in thermal equilibrium; in SdS spacetime the de Sitter space itself containing the black hole is compact and acts as a bounded box, implying thermal equilibrium. On the other hand, our results disagree with other approaches that argue equilibrium cannot be reached, due either to the system inevitably evolving towards empty de Sitter space or to phase transitions \cite{Kubi,Medved}.

In Section \ref{sec:temp_SdS} we introduce our new temperature result for SdS spacetime, explaining its important properties and its behaviour under certain limits. The derivation of the temperature is then presented in full in Section \ref{Deriv}. An appendix showing the effectiveness of our topological method in defining the spacetime temperature in the Nariai limit, where both horizons coalesce, is included, as is another appendix showing how the same results emerge from a dimensional reduction from four to two spacetime dimensions.

\section{The temperature of Schwarzschild-de Sitter spacetime} \label{sec:temp_SdS}

In this work, each spacetime analysed will have Riemannian signature. This is important because our topological Hawking temperature formula takes a Riemannian, not pseudo-Riemannian, space and calculates a topological Hawking temperature from the space's Euler characteristic. The Euler characteristic can only be calculated if a Euclideanised geometry is used. However, the formula gives temperatures valid in pseudo-Riemannian spacetimes as one can perform a Wick rotation if necessary as a final step in any temperature calculations \cite{ourpaper2}. This work studies the four-dimensional Euclidean de Sitter space and four-dimensional Euclidean Schwarzschild-de Sitter space which will be denoted by $\rm{EdS}_{4}$ and $\rm{ESdS}_{4}$ respectively below.

The $\rm{ESdS}_{4}$ metric describing a Schwarzschild black hole in de Sitter space is given by \cite{Bousso1}:
\begin{equation} \label{SdS_met}
\begin{split}
ds^2 =&\left( 1 - \frac{2M}{r} - \frac{\Lambda}{3}r^2 \right)d\tau^2 + \frac{1}{\left( 1 - \frac{2M}{r} - \frac{\Lambda}{3}r^2 \right)}dr^2 \\
&+ r^2 d\theta^2 + r^2 \mathrm{sin}^2 \theta d\phi^2,
\end{split}
\end{equation}
where $M$ is the black hole's mass and $\Lambda$ is the (positive) cosmological constant, while $r,\theta,\phi$ are the usual radial and angular variables.

As the new global Hawking temperature of the $\rm{SdS}_{4}$ spacetime we find:
\begin{equation} \label{SdS_temp}
T_{\rm{H}}=\frac{\left( r_{c}^3 - r_{b}^3 \right)\left( r_{b}^3 r_{c}^3 \Lambda^2 + 18M^2 \right)}{36 r_{b}^3 r_{c}^3 \pi},
\end{equation}
where $r_{c}$ and $r_{b}$ is the position of the cosmological and black hole horizon respectively in the coordinate system of metric (\ref{SdS_met}). We present the formula initially here before showing its full derivation in Section \ref{Deriv} in order to immediately explain its important properties and behaviours in certain limits.

\subsection{Interesting properties}

We now look at our temperature result (\ref{SdS_temp}) closely to show that it makes physical sense and matches intuition. Importantly, the temperature is always positive as $r_{c}$ must always be greater than $r_{b}$. The temperature formula also contains both the black hole's mass and the spacetime's cosmological constant as one would expect. It can be seen that (\ref{SdS_temp}) vanishes in the limit $r_{b} \rightarrow r_{c}$ (known as the Nariai limit) but this is acceptable because our starting metric (\ref{SdS_met}) becomes degenerate and so cannot be used in this case. The ``Nariai metric" should be used in this limit, giving a known, nonzero Hawking temperature which we will derive using our method in Appendix A. One key feature of our Hawking temperature result is that, despite its complicated form, it reduces to known horizon temperatures in both of the limits in which only one horizon remains. We will show this limiting behaviour below, providing evidence of the temperature result's reliability.

\subsection{The Schwarzschild limit}

In the Schwarzschild limit, i.e. when the cosmological constant becomes zero, leaving only a black hole horizon, one can see that the metric (\ref{SdS_met}) reduces to that describing the Schwarzschild black hole. The temperature (\ref{SdS_temp}) must reduce to the known Schwarzschild black hole Hawking temperature in this limit, which it indeed does.

The SdS spacetime temperature (\ref{SdS_temp}) with the Euler characteristic $\chi$ of the space left as a free variable (this feature will be explained fully in Section \ref{Deriv}) is:
\begin{equation} \label{SdS_temp_with_chi}
T_{\rm{H}}=\frac{\left( r_{c}^3 - r_{b}^3 \right)\left( r_{b}^3 r_{c}^3 \Lambda^2 + 18M^2 \right)}{9 r_{b}^3 r_{c}^3 \pi \chi}.
\end{equation}
The reason $\chi$ is left undetermined here is because the Schwarzschild limit involves a violent shift in topology. The transition from $\rm{ESdS}_{4}$ to Euclidean Schwarzschild space leads to a shift in topology from $S^2 \times S^2$ to $\mathbb{R}^2 \times S^2$ when $\Lambda$ reaches zero \cite{Garriga,Bousso2}. This transition manifests as a halving of the value of the Euler characteristic, from $\chi=4$ to $\chi=2$. Clearly, when $\chi=4$, temperature (\ref{SdS_temp_with_chi}) becomes (\ref{SdS_temp}).

Now we can see that the temperature (\ref{SdS_temp_with_chi}) reduces to the Schwarzschild black hole temperature in the appropriate limits. The limits to be taken are: firstly, $\chi \rightarrow 2$ as explained above, secondly $\Lambda \rightarrow 0$, then $r_{c} \rightarrow \infty$. Finally, $r_{b}\rightarrow 2M$ gives the known Schwarzschild black hole temperature $T_{\rm{H}}=(8\pi M)^{-1}$.

\subsection{The de Sitter limit}

The opposite limit to that studied above is where the black hole mass reduces to zero and only the cosmological horizon is left, leaving pure de Sitter space. The $\rm{EdS}_{4}$ space is described by the metric \cite{Bousso1}:
\begin{equation} \label{dS_met}
\begin{split}
ds^2 =&\left( 1 - \frac{\Lambda}{3}r^2 \right)d\tau^2 + \frac{1}{\left( 1 - \frac{\Lambda}{3}r^2 \right)}dr^2 \\
&+ r^2 d\theta^2 + r^2 \mathrm{sin}^2 \theta d\phi^2,
\end{split}
\end{equation}
and has the topology of the four-sphere $S^4$ (n.b. the topology of the spatial slices of de Sitter space are subtle to study and so much care is needed, see for example the discussion in Ong and Yeom \cite{Ong1}; our work here takes into account only the global topology of the space and so these difficulties are completely circumvented when calculating thermodynamical quantities).

Starting from (\ref{SdS_temp_with_chi}) one sets $\chi=2$ (as in the de Sitter limit the space's topology changes from $S^2 \times S^2$ to $S^4$), then letting $M \rightarrow 0$, $r_{b} \rightarrow 0$ and $r_{c} \rightarrow \sqrt{3/\Lambda}$ collapses the topological temperature (\ref{SdS_temp_with_chi}) to the known de Sitter cosmological horizon temperature \cite{Bousso1}:
\begin{equation}
T_{\rm{H}}=\frac{1}{2\pi}\sqrt{\frac{\Lambda}{3}}.
\end{equation}

\section{The derivation from topology} \label{Deriv}

The situation of de Sitter space containing a Schwarzschild black hole (SdS spacetime) is interesting and has several subtleties. Due to the presence of two horizons in the spacetime, a black hole event horizon and a cosmological horizon, and the absence of a flat asymptotic limit, it is unclear how horizon temperatures should be defined. In the words of a recent publication ``The question of the proper temperature for [the Schwarzschild-de Sitter spacetime] is still debated in the literature" \cite{Pappas}. In this recent work six different temperatures were presented: one associated only to the black hole horizon, one purely to the cosmological horizon, both defined from their surface gravities, and four attempting to describe a sensible effective temperature for the whole spacetime taking both emitting horizons into account. As we will outline below, using our topological approach a Hawking temperature can be derived for the whole spacetime. This temperature is not only different to previous suggestions but we believe a more natural way of defining a \emph{global} temperature for SdS spacetime than previously found. In this section we derive our new SdS spacetime temperature (\ref{SdS_temp}) using topological methods, based on the Euler characteristic of four-dimensional spaces, building on previous work by the authors studying the thermodynamics of simpler black hole systems using the same method.

Recently, the authors derived black hole Hawking temperatures by calculating the Euler characteristic of compact spaces \emph{with boundary}, these spaces being Euclideanised versions of spacetimes containing the black holes \cite{ourpaper2}. The expression for the Euler characteristic of each space can be rearranged, using the technique of integrating over Euclidean time with period of inverse temperature, to give a simple expression for each black hole's Hawking temperature \cite{Pad}.

Previously, the authors studied only 2d spacetimes (or dimensionally-reduced 4d spacetimes), whereas in this work a full 4d treatment of the SdS spacetime is presented. We now lay out the mathematical basis for the work before deriving our thermal equilibrium temperature (\ref{SdS_temp}).

The Euler characteristic of an $n$-dimensional compact manifold $M^n$, which later we identify as the Euclideanised black hole spacetime manifold, can be defined as the integral of a density form $\Omega$ over the manifold: $\chi=\int_{M^n}\Omega$. This density $\Omega$ is necessarily given, in many treatments, in the guise of differential forms, however all calculations carried out in this work will employ the density in Riemannian coordinates as we wish to study black hole metrics in particular coordinate systems \cite{Morgan,Chern1,Chern2}.

Chern proved that $\Omega$, originally defined in $M^n$, can be defined in a larger manifold $M^{2n-1}$, itself formed by the unit vectors of $M^{n}$ \cite{Chern1,Chern2}. The form $\Omega$ is equal to the exterior derivative of another form, $\Pi$, of degree $n-1$ which is defined in $M^{2n-1}$ via $\Omega=-d\Pi$. The integral of $\Omega$ over $M^n$ is equal to the same integral over a submanifold $V^n$ of $M^{2n-1}$, and by Stokes' theorem is also equal to the integral of $\Pi$ over the boundary of $V^n$.

A manifold \emph{with boundary} requires an important correction to the value of $\chi$, becoming \cite{Gibbons}:
\begin{equation} \label{eq:difference}
\chi=\int_{\partial V}\Pi-\int_{\partial M}\Pi.
\end{equation}
The submanifold $V^n$ of $M^{2n-1}$ is crucial as its boundaries are defined to be the fixed points (the zeros) of the unit vector field defined in $M^n$. Strikingly, it is known that any unit vector field in $M^n$ can be chosen in order to find the value of $\chi$ for that manifold; the most natural choice when studying black hole theory (adopted in this work) is to choose a Killing vector field for this field in $M^n$ \cite{Gibbons,Chern2,Brass}.

In our previous work studying different black hole spacetimes \cite{ourpaper2}, based on earlier research by Gibbons and Kallosh \cite{Gibbons}, an arbitrarily-positioned outer boundary to the Euclideanised spacetime had to be added in order to enforce its compactness, a requirement for defining the space's Euler characteristic. Importantly, in the case of $\rm{ESdS}_{4}$ space, the space is \emph{already compact} and without boundary. No artificial boundary, or ``cap", is needed. This requires a slightly different construction of the Euler characteristic than that used in our previous work, arguably a simpler one.

In the past, every Euclideanised black hole spacetime we studied required the above-mentioned cap on the radial coordinate of the space, as well as a boundary correction term in order to find that space's Euler characteristic -- as in formula (\ref{eq:difference}). The second term in (\ref{eq:difference}) is the boundary correction \cite{Gibbons,ourpaper2,Eguchi}. For the topologies studied in this work the boundary correction to the Euler characteristic is not appropriate as $\rm{ESdS}_{4}$ space is already compact and without boundary. Thus, the integral $\chi=\int_{\partial V}\Pi$ is sufficient.

To be explicit: $V$ denotes a space whose boundaries $\partial V$ are given by the fixed points or fixed surfaces of a Killing vector field on the Euclideanised black hole spacetime. In $\rm{ESdS}_{4}$ space there are two fixed-point surfaces (called ``bolts" in Hawking and Gibbons' elegant work \cite{Hawking}) of time isometry. In other words: the space has two Killing horizons (when the black hole mass obeys $0 < M < 1/(3\sqrt{\Lambda})$). Therefore, $V$ has two boundaries, leading to an expression for the Euler characteristic of $\rm{ESdS}_{4}$ space:
\begin{equation} \label{eq:compact_chi}
\chi=\int_{r_{c}}\Pi-\int_{r_{b}}\Pi.
\end{equation}

Before we derive (\ref{SdS_temp}) let us take a quick detour through two dimensions. For a 2d spacetime (or a dimensionally-reduced 4d spacetime) containing a black hole, its Euler characteristic (after Euclideanisation) can be rearranged to give a very simple, general formula for the Hawking temperature of the black hole \cite{ourpaper2}:
\begin{equation} \label{eq:main1}
T_{\mathrm{H}}=\frac{\hbar c}{4\pi \chi k_{\mathrm{B}}}\sum_{j \leq \chi}\int_{r_{\mathrm{H_{j}}}}\sqrt{g}Rdr,
\end{equation}
where $\hbar$ is the reduced Planck constant, $c$ is the speed of light in vacuum, $k_{\rm B}$ is Boltzmann's constant, $R(r)$ is the Ricci scalar which depends only on the `spatial' variable $r$, $g$ is the (Euclidean) metric determinant, $r_{\rm H_{j}}$ is the location of the $j$-th Killing horizon and $\chi$ is the Euler characteristic of the black hole's Euclidean `spacetime'. The symbol $\sum_{j \leq \chi}$ is a sum over all the Killing horizons, where one must pay attention to the sign of each term in the sum, which can be positive or negative depending on the specific features of each horizon -- the overall temperature result is however always positive.

Unfortunately, due to the greater complexity of 4d spacetimes as well as the presence of {\em two} different types of horizon studied in this work (the black hole and cosmological horizons), a simple 4d version of formula (\ref{eq:main1}) cannot be found. Despite this, the derivation of Hawking temperatures in 4d spacetimes is still simple to carry out using topology, as we will show below.

We now return to four dimensions, deriving $\chi$ for $\rm{ESdS}_{4}$ space, leading straight to its thermal equilibrium temperature. A general formula for the Euler characteristic of a four-dimensional Riemannian manifold given now in terms of local coordinates is:
\begin{equation} \label{chi_4d}
\chi=\frac{1}{32\pi^2}\int d^4 x \sqrt{g}\left( K_{1} - 4R_{ab}R^{ab} + R^2 \right),
\end{equation}
where $K_{1}\equiv R_{abcd}R^{abcd}$ is the Kretschmann invariant and $R_{ab}$ is the Ricci tensor. The $\rm{ESdS}_{4}$ metric (\ref{SdS_met}) has two Killing horizons, two fixed-point surfaces at $r_{b}$ and $r_{c}$, the black hole and cosmological horizon respectively, and so $\chi$ must be evaluated at both of these radial limits. Metric (\ref{SdS_met}) has geometry: $R=4\Lambda$, $R_{ab}R^{ab}=4\Lambda^2$ and $g=r^{4}\rm{sin}^2 (\theta)$. 

First, we must integrate (\ref{chi_4d}) over its angular coordinates, after substituting in the geometric values given above, yielding:
\begin{equation}
\chi=\frac{1}{8\pi}\int dt dr (r^2 K_{1}).
\end{equation}
One then integrates over Euclidean time, with period of inverse Hawking temperature $\beta$ \cite{Gibbons,Pad}. This is followed by inserting the Euler characteristic value for $\rm{ESdS}_{4}$ space $\chi=4$, introduced earlier, leading to:
\begin{equation}
1=\frac{\beta}{32\pi}\int dr (r^2 K_{1}).
\end{equation}
As explained earlier, this integral must be evaluated over both Killing horizons of $\rm{ESdS}_{4}$ space. After substituting in $\beta=(T_{\mathrm{H}})^{-1}$, this leaves:
\begin{equation} \label{eq:sign_illum}
T_{\rm{H}}=-\frac{1}{32\pi}\left( \int_{r_{c}}r^2 K_{1}dr - \int_{r_{b}}r^2 K_{1}dr\right).
\end{equation}

N.b. the signs before each integral in formula (\ref{eq:compact_chi}). The form $\Pi$ when put into local coordinates picks up a minus sign, hence the signs before each integral in expression (\ref{eq:sign_illum}) are correct and consistent with formula (\ref{eq:compact_chi}).

$\rm{ESdS}_{4}$ space (\ref{SdS_met}) has $K_{1}=(8\Lambda^2 /3) + (48M^2/r^6)$, inputting this value into (\ref{eq:sign_illum}) and evaluating both radial integrals, finally gives our form for the Hawking temperature of:
\begin{equation}
T_{\rm{H}}=\frac{\left( r_{c}^3 - r_{b}^3 \right)\left( r_{b}^3 r_{c}^3 \Lambda^2 + 18M^2 \right)}{36 r_{b}^3 r_{c}^3 \pi},
\end{equation}
which is a new result for the SdS spacetime temperature.

It is known that Hawking temperatures \emph{can be} defined separately for each horizon in SdS spacetime, derivable from each of their surface gravities in the standard way, and one finds a black hole temperature $T_{0}=(1-\Lambda r_{b}^2)/(4\pi r_{b})$ and cosmological horizon temperature $T_{c}=-(1-\Lambda r_{c}^2)/(4\pi r_{c})$ \cite{Pappas}. The new SdS spacetime temperature we find in this work (\ref{SdS_temp}) is the exact \emph{geometric average} of these two known temperatures, i.e. $T_{\mathrm{H}}=\left( T_{0}+T_{c} \right)/2$, as can be proven using analytical expressions for the horizon positions (given in Appendix B). Our approach confirms that there can be a thermal equilibrium in SdS spacetime as previously argued \cite {Shanki} and that it exists at our new derived temperature $T_{\mathrm{H}}$.

We believe that the SdS spacetime equilibrium temperature found in this work is more natural than others previously suggested in the literature. We make no thermodynamical assumptions as in past published temperature calculations, and another advantage of the topological approach is that an asymptotically flat limit is not crucial to the derivation of the temperature, with only the fixed-point behaviour of the Killing vector fields on the spacetime important.

It is hoped that in future research this thermal equilibrium will be studied further from a topological standpoint as outlined here.

\section{Conclusions}
We have introduced the first topological calculation of the global Hawking temperature of the Schwarzschild-de Sitter spacetime, based on the careful computation of the Euler characteristic. We have checked that our new formula (\ref{SdS_temp}) collapses to known expressions in the Schwarzschild limit and the de Sitter limit. The Nariai limit is also treated in Appendix A. Our topological approach is only applicable to find equilibrium temperatures; we deduce that the Schwarzschild black hole in the de Sitter spacetime can indeed reach an equilibrium temperature, which is the arithmetic average of the two horizons' Hawking temperatures.

\section*{Acknowledgments}

F.B. acknowledges support from EPSRC (UK) and the Max Planck Society for the Advancement of Science (Germany). F.B. thanks Prof Yen Chin Ong for useful email exchanges. L.D.M.V. acknowledges support from EPSRC (UK, Grant No. EP/L015110/1) under the auspices of the Scottish Centre for Doctoral Training in Condensed Matter Physics.

\section*{Appendix A}

In the Nariai limit where the black hole mass reaches $M \rightarrow 1/(3\sqrt{\Lambda})$ the two horizons meet, $r_{b} \rightarrow r_{c}$, and $r_{b}=r_{c}=1/\sqrt{\Lambda}$. Metric (\ref{SdS_met}) is no longer effective in this limit as it becomes degenerate. Instead, the Euclidean Nariai metric can be used \cite{Eune}:
\begin{equation} \label{Nariai_met}
ds^2=\frac{1}{\Lambda}\left( \mathrm{sin}^2 \xi d\tau^2 + d\xi^2 +  d\theta^2 + \mathrm{sin}^2 \theta d\phi^2 \right).
\end{equation}
This space has topology $S^2 \times S^2$ and therefore $\chi=4$. In these coordinates, the black hole horizon and cosmological horizon, the two bolts, are located at $\xi=0$ and $\xi=\pi$ respectively.

From the metric (\ref{Nariai_met}) one can calculate $K_{1}=8\Lambda^2$, $R=4\Lambda$ and $R_{ab}R^{ab}=4\Lambda^2$. Following the same topological scheme as shown earlier, the known Nariai black hole temperature emerges:
\begin{equation}
T_{\rm{H}}=\frac{1}{2\pi}.
\end{equation}
There are in fact two different Nariai black hole temperatures discussed in the literature. However, one of these has been shown to be a Tolman temperature \cite{Eune}, which changes with spacetime curvature. The other temperature discussed is the true Hawking temperature, and matches the one we find above using our topology approach, further showing the reliability and universality of our method.

\section*{Appendix B}

An interesting feature of the topological method of finding Hawking temperatures is that a dimensional reduction down to two dimensions is always possible. The dimensional reduction required consists of simply keeping the radial and temporal part of a Schwarzschild-like metric describing a black hole system, and cutting off the rest. This straightforward reduction then leaves a 2d metric describing the 4d spacetime, remarkably keeping all of the information required to find the correct Hawking temperature of the 4d system. An example of this calculation for several black holes, including the Kerr and charged hole, is given in our recent work \cite{ourpaper2}. A topological study of 4d and 2d descriptions of the Schwarzschild-AdS family of black holes, describing how a dimensional reduction can remove phase transition information, simplifying Hawking temperature calculations, will be presented in a future work.

The effectiveness of two-dimensional topological calculations of four-dimensional black hole Hawking temperatures, even in very complicated systems, suggests a kind of lower-dimensional information storage, redolent of the known area scaling of entropy and recent work in holography. This feature will be investigated in future work.

A dimensional reduction of Schwarzschild-de Sitter space will now be carried out, as will a 2d topological analysis, leading again to our new temperature result (\ref{SdS_temp}). The 4d metric (\ref{SdS_met}) is dimensionally-reduced to an effective 2d metric by simply removing the angular parts of the line element, leaving:
\begin{equation}
ds^2 =\left( 1 - \frac{2M}{r} - \frac{\Lambda}{3}r^2 \right)d\tau^2 + \frac{1}{\left( 1 - \frac{2M}{r} - \frac{\Lambda}{3}r^2 \right)}dr^2.
\end{equation}
A general formula for the Euler characteristic of a 2d space given in terms of local coordinates is:
\begin{equation} \label{chi_2d}
\chi=\frac{1}{4\pi}\int d^2 x \sqrt{g} R ,
\end{equation}
where here $R=(2\Lambda /3) + (4M/r^{3})$.
After the dimensional reduction, the Killing horizons become \emph{fixed points} of the metric, still located at $r_{b}$ and $r_{c}$. Therefore, the Euler characteristic must be set to $\chi=2$ (as opposed to $\chi=4$ for the 4d metric studied earlier).

As mentioned in the main text, horizon temperatures in 2d spacetimes can be found in every case using our Hawking temperature formula (\ref{eq:main1}), derivable from (\ref{chi_2d}); see \cite{ourpaper2} for details. For dimensionally-reduced SdS spacetime, this formula directly gives a thermal equilibrium temperature:
\begin{equation}
T_{\mathrm{H}}=\frac{\left( r_{c}-r_{b} \right) \Lambda + 3M\left( \frac{1}{r_{b}^2} - \frac{1}{r_{c}^2} \right)}{12\pi}.
\end{equation}
Despite appearances, this temperature is the exact same result as that found from the 4d topological analysis (\ref{SdS_temp}). This equality can be proven using known analytical expressions for the horizon positions. In the interval $0 < M < 1/(3\sqrt{\Lambda})$ (in which two real horizons exist), the horizon positions are located at $r_{b}=(2/\sqrt{\Lambda})\mathrm{cos}\left[ (\pi + \psi)/3 \right]$ and $r_{c}=(2/\sqrt{\Lambda})\mathrm{cos}\left[ (\pi - \psi)/3 \right]$, where $\psi=\mathrm{cos}^{-1}\left[ 3M\sqrt{\Lambda} \right]$ \cite{Rahman}.

\end{document}